# Magnetic domain-wall motion twisted by nanoscale probe-induced spin transfer

J. Wang[1#], L. S. Xie[1#], C. S. Wang[1], H. Z. Zhang[2], L. Shu[3], J. Bai[4], Y. S. Chai[2], X. Zhao[1], J. C. Nie[1], C. B. Cao[4], C. Z. Gu[2], C. M. Xiong[1*], Y. Sun[2], J. Shi[5], S. Salahuddin[6], K. Xia[1], C. W. Nan[3] and J. X. Zhang[1*]

1, Department of Physics, Beijing Normal University, Beijing, 100875, China.
2, Institute of Physics, Chinese Academy of Sciences, 100190, China.
3, School of Materials Science and Engineering, Tsinghua University, Beijing, 100084, China.
4, Research Center of Materials Science, Beijing Institute of Technology, Beijing, 100081, China.
5, Department of Physics and Astronomy, University of California, Riverside, California, 92521, USA.
6, Department of Electrical Engineering and Computer Sciences, University of California, Berkeley, California, 94720, USA

[#] These authors contributed equally to this work
Email: cmxiong@bnu.edu.cn and jxzhang@bnu.edu.cn

**A method for deterministic control of the magnetic order parameter using an electrical stimulus is highly desired for the new generation of spintronic and magnetoelectronic devices. Much effort has been focused on magnetic domain-wall motion manipulated by a successive injection of spin-polarized current into a magnetic nanostructure. However, an integrant high-threshold current density of $10^7 \sim 10^8$ A/cm$^2$ inhibits the integration of those nanostructures with low-energy-cost technology. In addition, a precise determination of the location of domain walls at nanoscale seems difficult in artificially manufactured nanostructures. Here we report an approach to manipulate a single magnetic domain wall with a perpendicular anisotropy in a manganite/dielectric/metal capacitor using a probe-induced spin displacement. A spin angular momentum transfer torque occurs in the strongly correlated manganite film during the spin injection into the capacitor from the nanoscale magnetized tip with an ultralow voltage of 0.1 V, where the threshold spin-polarized current density is ~$10^4$ A/cm$^2$ at the tip/manganite interface. The probe-voltage-controlled domain wall motion in the capacitor demonstrates a critical framework for the fundamental understanding of the manipulation of the nano-magnet systems with low energy consumption.**

Electrical control of magnetic spin structures at nanoscale is highly desired because of future demands for next-generation spintronics-based memory or logic devices [1-4]. This provides a strong motivation to explore materials or structures in which the spin degree of freedom can be manipulated using an electrical knob, particularly at room temperature. Recently, a variety of experimental and theoretical pioneering studies have proposed multiple concepts to achieve the above goals, including magnetoelectric coupling in multiferroics [5-7], magnetic anisotropy controlled by an interfacial charge density [8-11] or chemistry [12], and electric and magnetic field co-mediated domain-wall (DW) dynamics [8,9]. In addition, the magnetic DW motion driven by a spin-polarized current has attracted intense study in fundamental physics and in potential applications such as magnetic registers or the building blocks of racetrack memory [1-3,13,14]. However, the DC current input (with threshold values as high as $10^7$-$10^8$ A/cm$^2$) for spin-transfer torque (STT) [1,2,15-20] is a main factor in inducing thermal fluctuations as well as in power consumption [2,21]. Additionally, the deterministic control of the DW position in a magnetic nanostructure is crucial to the design of future nano applications [1,2,22,23].

Compared to the above STT strategy to drive DW motion by a DC current, the scanning-probe-based technique provides a platform to control the magnetic spin structure at nanoscale and to explore low-dimensional magnetism. Heinrich *et al.* at IBM demonstrated that a spin-polarized tunneling current in a scanning tunneling microscope system can be used to manipulate the spin flip in a single Mn atom [24]. Recently, Stapelfeldt *et al.* also theorized that a single magnetic domain wall can be moved by a spin-polarized current injected from a magnetized tip using a scanning tunneling microscope [25]. Nevertheless, such scanning-tunneling-microscope- based methods usually require extreme conditions such as low temperature and high vacuum [24], which are technically impractical and restrict possible applications. Here, we report a solution to deterministically manipulate the DW motion at a low energy cost using a nanoscale voltage probe under ambient conditions.

We demonstrate that a spin displacement across the interface of a magnetized tip and a magnetic film can drive the magnetic DW motion in a manganite/dielectric/metal capacitor. To generate a continuous spin displacement, an AC probe voltage (as low as 0.1 $V_{pp}$) is used to exert a local spin angular momentum transfer torque, where the instantaneous threshold spin-polarized current density is ~$10^4$ A/cm$^2$ at the contact interface of the magnetized tip and a ferromagnetic La$_{0.67}$Sr$_{0.33}$MnO$_3$ (LSMO) film. The DW motions and dynamics are analyzed using the Landau-Lifshitz-Gilbert-STT (LLG-STT) method. Our ambient control of the DW motion at nanoscale provides an important opportunity to study and manipulate nano-magnet systems.

In order to fulfill a spin displacement and angular momentum transfer torque at the interface of a downward-magnetized (red thick arrow) tip and a ferromagnetic thin film, contact geometry assisted with a scanning probe microscope (SPM) under ambient conditions is proposed as described in Fig. 1(a). A capacitor comprising ferromagnet/dielectric/metal layers is used as our model structure, which can be used to displace spins during electrical charging and discharging of the capacitor. Fig. 1(b) shows both Slonczewski and field-like torques between incident spin-polarized electrons and magnetizations when the nanoscale probe is on top of a domain wall of the ferromagnetic layer.

With the configuration in Fig. 1(a), a self-organized domain structure with a perpendicular anisotropy is desired to study the above proposed DW motion without any intricate processes (such as artificial notches or stochastic effects) [26,27]. LSMO has been widely investigated as a strongly correlated material [28-31], providing a good framework to study emerging phenomena related to the underlying coupling among lattice, spin, charge and orbital degrees of freedom [29,30]. Furthermore, well-ordered magnetic domain patterns with a perpendicular anisotropy can be fabricated by controlling the epitaxial growth [27], which can also be controlled by multiple macroscopic stimuli such as magnetic fields [26], thermal activation [32] and stress [26]. Therefore, LSMO-based capacitor provides a possible opportunity to probe the DW motion and dynamic using a local STT strategy.

LSMO thin films (~100 nm) with a magnetization of ~300 emu/cm$^3$ and a coercive field of ~50 Oe were grown on dielectric LaAlO$_3$ (LAO) substrates using pulsed-laser deposition (PLD), forming ferromagnetic/dielectric heterostructures. A distinct stripe-like magnetic domain pattern (100~120 nm in width) is shown in the magnetic force microscopy (MFM) image in Fig. 2(a), indicating a strong perpendicular magnetic anisotropy. The cross-sectional spin structure [33] of the domain wall in LSMO films is shown in Fig. 2(b), which is simulated using a 3D OOMMF method.

To study the probe-based electrical control of the magnetic DW motion in LSMO/LAO, a metal electrode (conductive silver) was coated on the backside of the dielectric LAO to form the capacitor, as shown in Fig. 2(c). In order to generate a continuous displacement of spin-polarized electrons at the tip/film interface, an AC voltage (0.1 V, 20 kHz) was applied on a downward-magnetized tip (Co-Cr coating) when it was scanning across the ferromagnetic layer with a contact mode. After the biased tip scanned the whole area, the stripes marked by the colored lines move away (Figs. 2(d)-2(e)) from the reference location labeled inside the top-left blue box. Figs. 2(c)-2(e) provides direct evidence that the domain walls creep (dark blue arrow) with the motion (purple arrows in Figs. 2(d)-2(e)) of the downward-magnetized tip with an AC voltage. This room-temperature observation provides a first glimpse of a magnetic DW motion controlled by STT in a correlated oxide at nanoscale, which can be achieved under ambient conditions through a scanning-probe-based technique.

With the above controls of the magnetic DW motion, we now turn to the manipulation of a single domain wall to understand its mechanism behind the AC stimulus. We applied the AC voltage on the tip using a point contact mode without scanning. As shown in the schematic of Fig. 3(a), the downward-magnetized tip was first placed on a domain wall with downward magnetization on the left and upward magnetization on the right. When an AC voltage (0.1 $V_{pp}$, 20 kHz) was applied on the still tip (location marked by green dots), the domain wall beneath the tip moved to the right as indicated by yellow arrow in Fig. 3(b). In order to understand the DW motion under the AC stimulus, the first 1/4 section of only one period of the AC voltage was applied on the still tip as seen in Fig. 3(c), where the spin-polarized electrons only flow from the downward-magnetized tip into the film. As we can observe in Fig. 3(d), the domain wall beneath the tip moved to the right (marked by yellow arrow) after the application of the fist 1/4 period of AC voltage, consequently the tip was located on top of the domain. When the tip with an AC voltage was placed on top of a domain (Fig. 3(e)), we observed no DW motion as seen in Fig. 3(f). Namely, the DW motion occurs at the first 1/4 period and there will be negligible DW motion even the spin-polarized electrons keep displacing across the tip/film interface afterwards. These results comply with our proposed DW motion driven by nanoscale-probe-induced STT (Fig. 1), which is only effective when the tip is on top of the domain wall.

To provide a quantitative understanding of this probe-induced STT and DW motion in the LSMO-based capacitor, we analyzed the instantaneous threshold spin-polarized current density at the tip/film interface. The application of the AC voltage on a capacitor can be regarded as a continuous charging/discharging process [34], in which the total electrons flow across the contact interface (the contact area is determined by the tip radius) of the downward-magnetized tip and film surface. In one charging process, the total polarized spins (S) displaced across the tip/film interface can be expressed as

$$S = \varepsilon_0 \varepsilon \frac{A}{ed} pV \qquad (1)$$

where $A$, $e$, $d$, $\varepsilon_0$ and $\varepsilon$ are the area of the planar electrode, electron charge, thickness, vacuum permittivity and the relative permittivity of the capacitor, respectively. $V$ is the applied probe voltage, and $p$ is the spin polarization of the magnetic tip. The capacitance ($\varepsilon_0 \varepsilon \frac{A}{d}$) in our probe-based capacitor remains constant at ~4 pF from 100 Hz to 100 kHz, measured using an LCR meter combined with the SPM. In order to obtain the threshold current density at the interface, we first studied the areal fraction ($fr$) of the domain walls parallel to the tip scanning direction as a function of AC voltage. A threshold probe voltage of 0.1 $V_{pp}$ was obtained at a frequency of 20 kHz. Considering the contact area of the downward-magnetized tip and the LSMO film (contact radius less than 5 nm [35,36]), the maximum spin-polarized charge density across the interface in one dynamic charging process can reach ~0.5 C/cm$^2$ at a voltage of 0.1 $V_{pp}$. Within $5 \times 10^{-5}$ seconds (an AC frequency of 20 kHz), an instantaneous threshold current density was on the order of ~$10^4$ A/cm$^2$ during the spin displacement at the tip/film interface.

In addition, we have also analyzed the threshold current density with a quasi-static spin-polarized current. Similar to the case of a ferromagnet with a narrow domain wall (~10 nm) [33,37], the threshold current density for DW motion in LSMO can be estimated as [38,39]

$$J_c = \frac{2H_c\mu_B}{ena^3R_wA} \quad (2)$$

where $\mu_B$ is the Bohr magneton, and $e$ is the electron charge. $H_c$ is the depinning field of the domain walls, which is on the order of $10^{-2}$ T in our case [33]. The carrier density per unit cell is $na^3 = 1.6$, where $n$ is the charge-carrier concentration and $a$ is the lattice constant of the magnetic unit cell [40]. The DW area resistance is $R_wA = 1.9 \times 10^{-15}$ $\Omega m^2$ [33], where $R_w$ is the wall resistance and $A$ is the cross-sectional area. Therefore, the calculated threshold $J_c$ is approximately $3.8 \times 10^4$ A/cm$^2$ in the LSMO film with a perpendicular anisotropy. This is in good agreement with the experimental threshold spin-polarized current density of ~$10^4$ A/cm$^2$ during one charging process across the tip/LSMO interface.

To further elucidate the dynamics of the DW motions with a current-perpendicular-to-plane (CPP) geometry driven by the tip-induced quasi-static spin-polarized current, we use the LLG-STT equation [41]

$$\frac{d\boldsymbol{m}}{dt} = -|\gamma|\boldsymbol{H}_{eff} \times \boldsymbol{m} + \alpha \boldsymbol{m} \times \frac{d\boldsymbol{m}}{dt} + u \cdot \boldsymbol{m} \times (\boldsymbol{m} \times \boldsymbol{p}) + \beta \cdot u \cdot \boldsymbol{m} \times \boldsymbol{p} \quad (3)$$

The unit vectors $\boldsymbol{m}$ and $\boldsymbol{p}$ denote the local magnetization direction and the polarization of the current, $\gamma$ is the gyromagnetic constant, $H_{eff}$ denotes the micromagnetic effective field, and $\alpha$ is the Gilbert damping constant. The last two terms on the right describe Slonczewski and field-like torques on $\boldsymbol{m}$, where $u = JPg\mu_B/(2eM_s)$, with $J$ (the current density) and $P$ (polarization rate); for LSMO, the factor $g\mu_B/(2eM_s)$ can be $2 \times 10^{-11}$ m$^3$/C. $\beta$ is a dimensionless coefficient that indicates the ratio between the field-like (out-of-plane) and the Slonczewski (in-plane) torques. The dynamic process of the DW motions was simulated using a 3D OOMMF method with a sample size of 180 ×12 × 60 nm$^3$ using a cubic cell of 3 × 3 × 3 nm$^3$. The parameters, such as the effective room-temperature magnetic anisotropy energy ($K = 1.5 \times 10^4$ J/m$^3$) and the exchange stiffness constant ($A_{ex} = 1.94 \times 10^{-12}$ J/m) for the simulation were obtained from a previous study [33]. In the dynamic simulation, the equivalent AC current density at the interface was controlled with the electrical input as illustrated in Fig. 4(a), where the maximum current density is $10^4 A/cm^2$ in one period. Figure 4(b) depicts the instantaneous distribution of the spin-polarized current density (stage Ⅲ in Fig. 4(a)) at the cross-section of the ~100-nm-thick LSMO film beneath the magnetized tip. The dynamics of the DW motions for the AC spin-polarized current ($J \propto (r + R)^{-2}$) in the CPP geometry are shown in Fig. 4(c). The downward (red arrow in stage I) and upward (yellow arrow in stage I) spins are separated by a domain wall, as in the case of LSMO [33], where the blue and red pixels correspond to $-M_z$ and $+M_z$, respectively. During the initial state of one period of the AC current, the position of the downward-magnetized (dark blue arrows) tip is assumed to be on top of the wall (stage I). When the spin-polarized current density is lower than the threshold value, no noticeable change of the DW location (stage Ⅱ) is observed. When the current density reaches $10^4 A/cm^2$ (first 1/4 period), the domain wall completely moves out of the location of the tip (stage Ⅲ) and the tip was located on top of domain. Therefore, the reversed spin displacement beneath the tip in the second-half period will give rise to a negligible spin angular momentum transfer torque and DW motion (stage Ⅳ). In good agreement with our observations in Figs. 3(d) and 3(f), our simulation further demonstrates that the domain wall motion occurs within ~$10^{-7}$ seconds and completely moves away during the first half cycle ($2.5 \times 10^{-5}$ seconds) of the AC current in the first period. Although the spin displacement continues in the following electrical input, the spin angular momentum transfer torque does not occur when the tip located on top of domain.

Possible contributions from thermal fluctuation and stray fields of the downward-magnetized tip have been excluded by producing pulsed spin displacements across the tip/LSMO interface. One effective way to reduce the possible local thermal fluctuation is to use a pulsed spin flow into the capacitor in comparison to the continuous AC spin displacement, which also provides additional degrees of freedom to study the directions of the DW motions through a one-way spin injection/pumping. We combined the SPM system with a programmable waveform generator. To achieve the one-way spin injection/pumping across the tip/film interface, asymmetric triangular voltage waveforms were programmed. The instantaneous magnitudes of current density are ~$10^4$ A/cm$^2$ and ~10 A/cm$^2$ during the spin injection and pumping, so that the spin angular momentum transfer torque only occurs during the injection, or vise versa. By programming the scanning step width and pulse interval, we are also able to achieve that the one-way spin injection always occurs at the domain wall.DW motion was also observed using this triangular waveform.

Utilizing the benefits of a programmable spin displacement, we further studied the DW motions using different current directions, controlled by a one-way spin injection and pumping across the interface of the downward-magnetized tip (dark blue arrows in the bottom panels of Figs. 5(a) and 5(d)) and LSMO film. The tip was placed on top of the domain wall with downward magnetization on the left and upward magnetization on the right, as illustrated in the bottom panels of Figs. 5(a) and 5(d). When the current flow from the downward-magnetized tip into the film, we observed that the domain wall moved to the right (labeled by pink arrow in Fig. 5(b)) and the tip located on top of the domain with a downward magnetization due to the DW motion, as shown in Fig. 5(b); while when the current pumped from the film into the tip, the domain wall moved to the left (labeled by pink arrow in Fig. 5(e)) and the downward-magnetized tip located on top of the domain with an upward magnetization due to the DW motion, where the magnetization is opposite to that of the tip, as shown in Fig. 5(e). This phenomenon can be repeated and observed in multiple areas and the corresponding simulations for the DW motions during the spin injection and pumping can be seen in Figs. 5(c) and 5(f). Therefore, we conclude that the stray field from the magnetized tip does not play a significant role in DW motion.

The potential extrinsic factors such as the possible elevated temperature [42,43], stray fields from the Co-Cr-coated tip [35], and mechanical forces [44] were further studied using a non-magnetic Pt-coated tip with an AC voltage, Co-coated tip with a DC voltage, Co-coated tip without any electrical input, respectively. All the above tests demonstrate that magnetic domain pattern still remains after the tip scan across the sample in a contact mode, indicating negligible contributions on DW motion from thermal effect, electromagnetic induction, microwave excitation [45], static charge, mechanical force, and stray field, etc. Beside, we also estimated the distributions of the magnetic field and magnetostatic energy density beneath the Co-coated tip. The results indicate that the maximum stray field and magnetostatic energy density are ~0.5 Oe and $10^{-3}$ J/m$^3$, respectively. These values are much lower than the threshold field and energy density of the DW motion in LSMO [33,46].

To conclude, we demonstrate that a local spin displacement can induce a spin angular momentum transfer torque at the interface of the nanoscale magnetized tip and a strongly correlated oxide in the LSMO-based capacitor, giving rise to deterministic control of the motion of magnetic domain walls with a perpendicular anisotropy. The threshold spin-polarized current density of ~$10^4$ A/cm$^2$ at the tip/film interface demonstrates its ultralow energy consumption. We further utilized the LLG-STT equation to analyze the dynamic process of the magnetic DW motions induced by the spin displacement in this CPP geometry, which is in good agreement with our observations. This nanoscale electrical control of the magnetism in the LSMO-based capacitor configuration provides us with a brand-new platform to understand and manipulate the spin structures in strongly correlated systems with low energy consumption.

**Figures:**

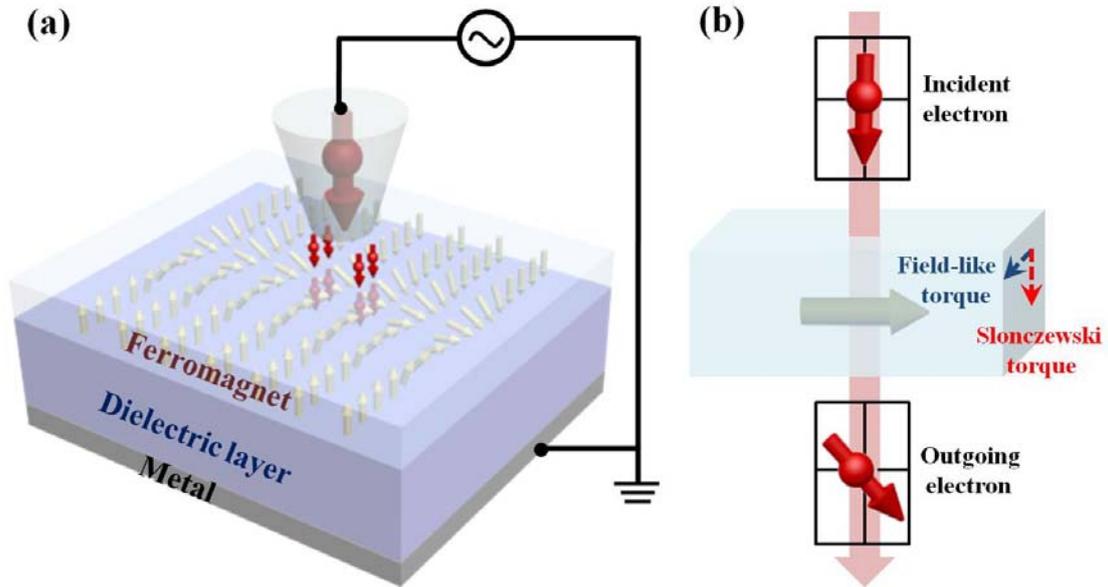

FIG. 1. Schematics of probe-induced STT in a ferromagnet-based capacitor. (a) Contact geometry of vertical spin displacement between a downward-magnetized (red thick arrow) tip and a ferromagnet/dielectric/metal capacitor induced by spin displacement, where the Bloch wall is used for example. (b) The interaction between the incident spin-polarized electron (red arrow) and the magnetization (gray arrow) of the DW in the ferromagnetic layer, both in-plane (Slonczewski) and out-of-plane (field-like) torques are marked by red and blue dotted arrows, respectively.

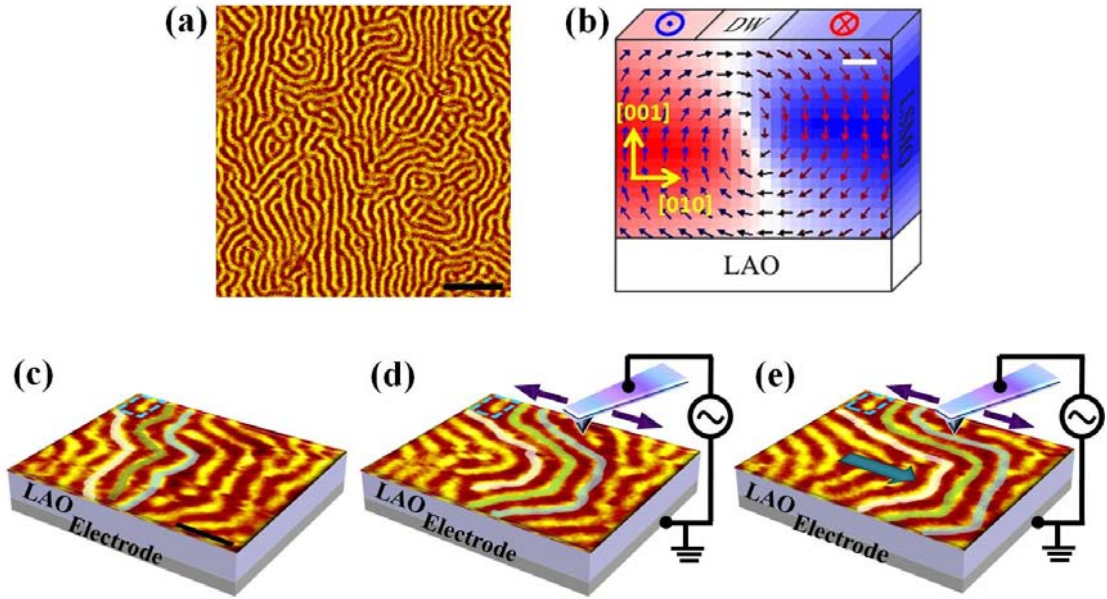

FIG. 2. Magnetic DW motions induced by nanoscale-probe-induced STT. (a) A top-view MFM image of an as-grown stripe-like magnetic domain in an epitaxial LSMO film. (b) A simulated cross-sectional spin structure of a single domain wall in LSMO film. (c) The as-grown stripe-like magnetic domain structure. (d) The domain structure when an AC voltage was applied to a scanning tip at the same area. (e) The magnetic domain when the AC voltage on the scanning tip was further applied at the same area, where the colored lines help indicate the movement (dark blue arrow) of the walls along the fast-scanning direction of the tip (purple arrows). Scale bar, 1μm (a), 10 nm (b) and 400 nm (c).

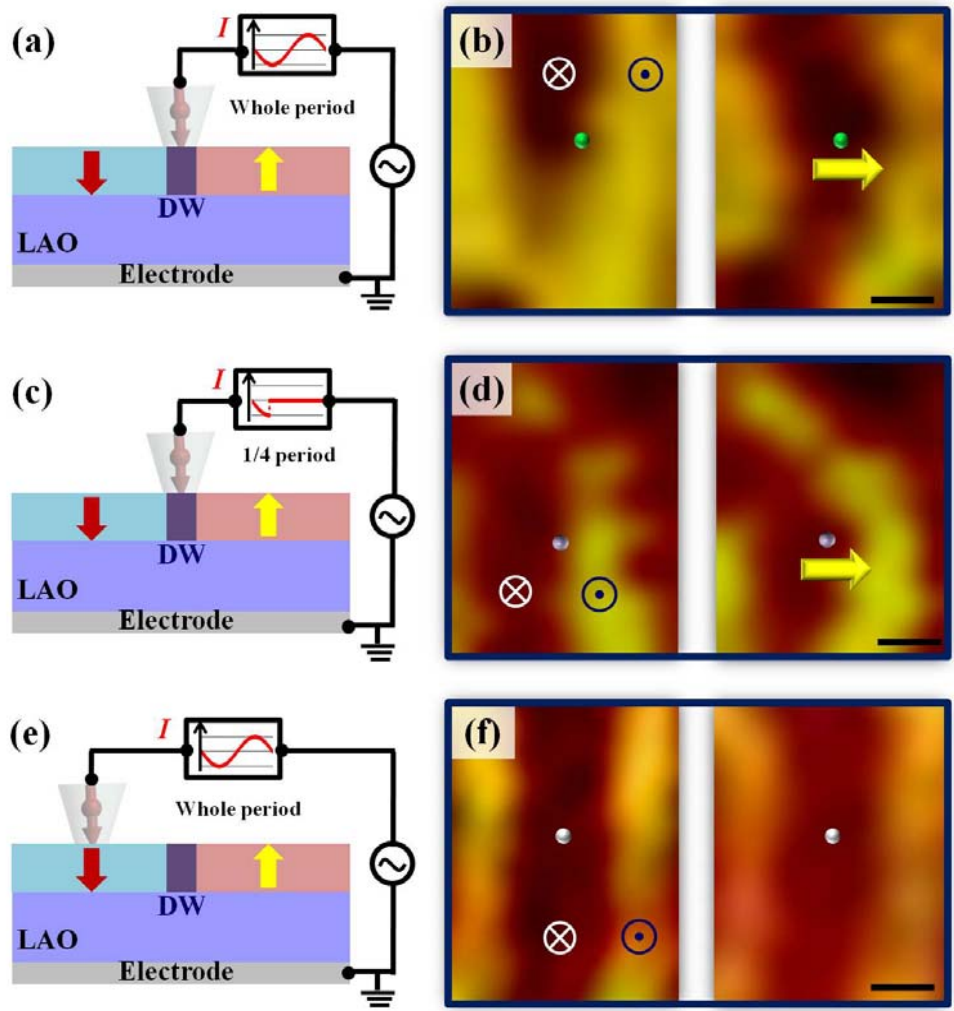

FIG. 3. Probe-induced single DW motion within an AC stimulus. Schematics of a point contact geometry with the tip placed on top of a domain wall (a,c) and domain (away from the wall) (e), where the tip was applied with a whole period (a,e) and 1/4 period (c) of the AC voltage, respectively. The waveforms in the black box in (a), (c) and (e) describe the instantaneous current input. (b,d,f) The corresponding MFM images before and after the still tip (location marked by color dots) was applied with an AC voltage, where we can clearly see that both DW motions occur (indicated by yellow arrows) when the tip was placed on top of a domain wall during either the whole period (b) or the first half cycle (d), but no DW motion occurs when the tip was placed on top of domain (f). (×) and (•) represent downward and upward magnetizations, respectively. Scale bar, 100 nm (b,d,f).

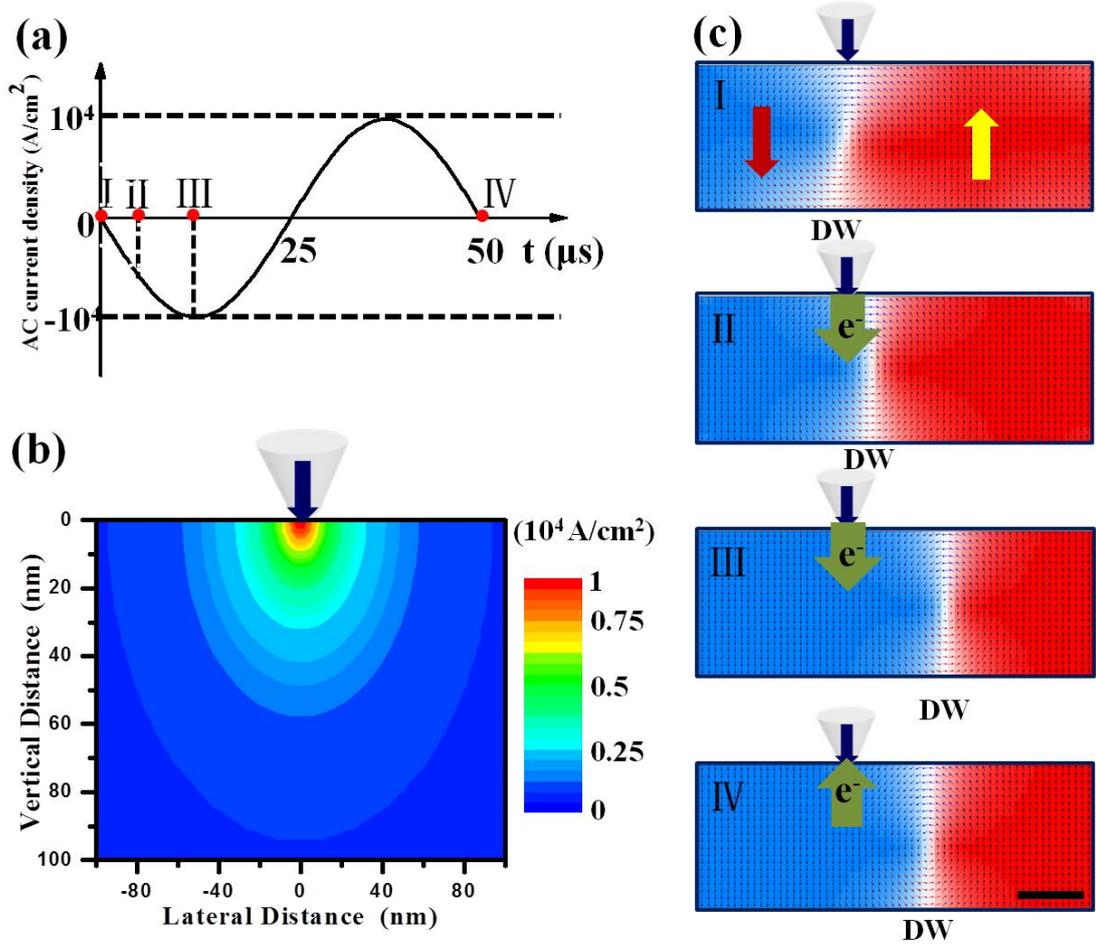

FIG. 4. Simulated dynamics of the DW motions during AC spin displacement. (a) The AC current density within one period used in the dynamic simulation, where I, II, III, and IV indicate different instantaneous states during one AC period. (b) The distribution of the instantaneous spin-polarized current density (stage III in (a)) at the cross-section of the ~100-nm-thick LSMO film beneath the downward- (dark blue arrow) magnetized tip. (c) The corresponding LLG-STT simulation of the dynamics of the DW motions under the AC stimulus, where I, II, and III indicate in detail that the DW completely moves out of the location of the tip with a downward magnetization (dark blue arrows) within the first-half cycle and IV indicates that the backward flow of spins during the second-half cycle exerts a negligible torque.

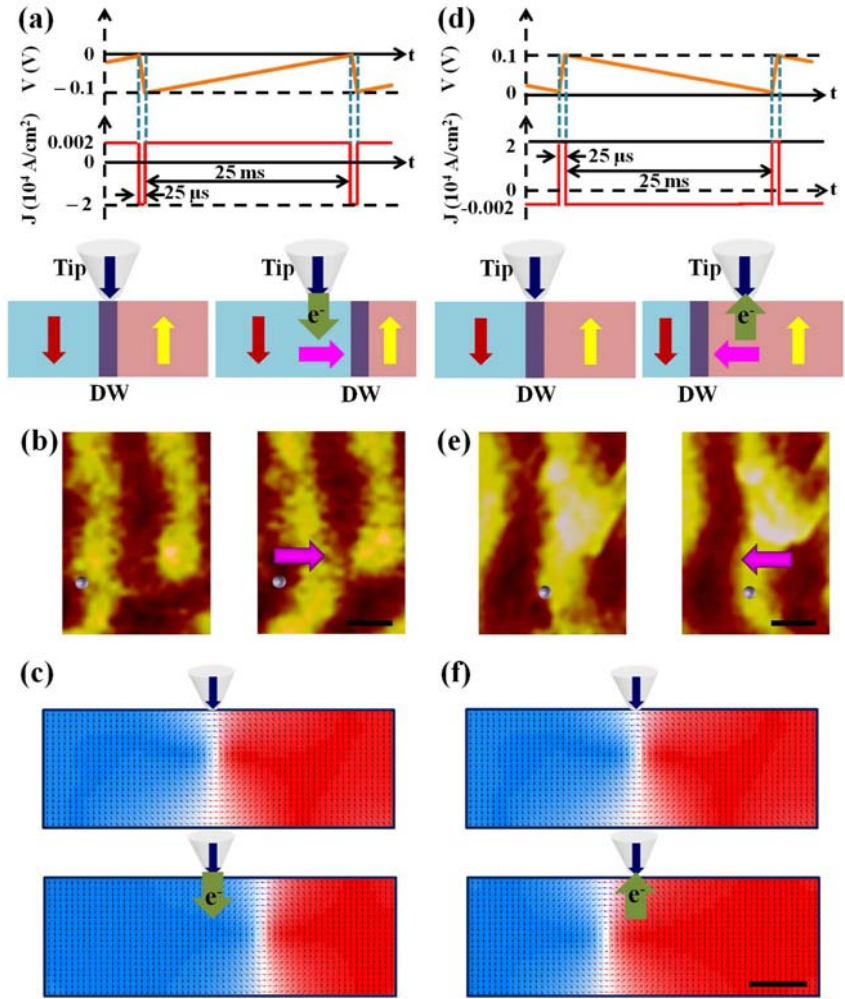

FIG. 5. Magnetic DW motions induced by one-way spin injection and pumping. The schematic diagrams (upper panels) of the triangular waveforms with the corresponding current densities and the schematics (bottom panels) of the DW motions using one-way spin injection (a) and pumping (d), where the dark blue arrows indicate the polarization direction of the downward-magnetized tip, the red and yellow arrows describe the spin directions of the magnetic domains, and the green arrows describe the flowing directions of the spins. The motions of the single domain wall separated by downward (left) and upward (right) magnetizations during one-way spin injection (b) and pumping (e), where the spin injection and pumping give rise to DW motions in an opposite way (purple dots describe the location of the tip, pink arrows indicate the direction of the DW motions). (c) and (f) show the corresponding simulations of the DW motions during one-way spin injection and pumping. Scale bar, 100 nm (b,e) and 30 nm (f).

19